\documentclass{PoS}

\usepackage{latexsym}
\usepackage{subfigure}
\usepackage{multicol}
\usepackage{multirow}
\usepackage{mathrsfs}
\usepackage{color}
\usepackage{verbatim}
\usepackage{amsmath,amsthm, amssymb}

\newcommand{\stat}{\textrm{stat}\,}
\newcommand{\kin}{\textrm{kin}\,}
\newcommand{\spin}{\textrm{spin}\,}
\newcommand{\bare}{\textrm{bare}\,}
\newcommand{\pole}{\textrm{pole}\,}

\newcommand{\match}{\textrm{match}\,}
\newcommand{\HQET}{\textrm{HQET}\,}
\newcommand{\QCD}{\textrm{QCD}\,}

\title{On one-loop corrections to matching conditions of Lattice HQET including $1/m_b$ terms}

\ShortTitle{On one-loop corrections to matching conditions of Lattice HQET including $1/m_b$ terms}

\author{
	\speaker{Piotr Korcyl} 	
\hfill{\footnotesize{\it DESY 13-204}}\\
        NIC, DESY, Platanenallee 6, 15738 Zeuthen, Germany\\
        E-mail: \email{piotr.korcyl@desy.de}}

\author{(for the ALPHA collaboration)}

\abstract{HQET is an effective theory for QCD with $N_f$ light quarks and a massive valence quark
if the mass of the latter is much bigger than $\Lambda_{\QCD}$. As any effective theory, HQET is
predictive only when a set of parameters has been determined through a process called matching.
The non-perturbative matching procedure including $1/m_b$ terms, developped by the ALPHA collaboration, consists of 19
carefully chosen observables which are precisely computable in lattice QCD as well as in lattice HQET.
The matching conditions are then a set of 19 equations which relate the QCD and HQET values of these
observables. We present a study of one-loop corrections to two generic matching observables involving correlation function
with an insertion of the $A_0$ operator. Our results enable us to quantify the quality of the relevant observables in 
view of the envisaged non-perturbative implementation of this matching procedure.
}

\FullConference{31st International Symposium on Lattice Field Theory - LATTICE2013,\\
		July 29 - August 3, 2013\\
		Mainz, Germany}

\begin{document}

In a problem involving a hierarchy of scales such as a lattice QCD simulation of heavy-light mesons one needs 
to employ an effective description of dynamics on one of the scales if lattices of affordable size are to be used. 
In a particular case of extraction of decay constants and form-factors of B-mesons, the ALPHA collaboration decided 
to use Heavy Quark Effective Theory \cite{eichten_hill} in order to account for the dynamics of the \textbf{b} quark.
A fully non-perturbative strategy was set up \cite{heitger_sommer,1,2} which consists of a non-perturbative matching 
step between HQET and QCD in a finite volume using the Schr\"odinger functional framework and of a non-perturbative 
evolution of HQET parameters using step scaling techniques up to volumes sufficiently large to perform full QCD calculations. 
The success of the matching step relies on a set of suitable QCD observables and their effective HQET counterparts 
which can be precisely evaluated in a Monte Carlo simulation. Apart of being precise, one also requires that the 
matching observables do not introduce artificially large $1/m_b^2$ corrections. The entire set of matching observables
was investigated at tree-level of perturbation theory in Ref.\cite{3} and the aim of this work is to 
report on the extention of that study to include one-loop corrections.

After introducing basic notation in section \ref{sec1} we describe two examples of matching observables in section \ref{sec2}
and discuss how to estimate the size of such unwanted $1/m_b^2$ corrections using lattice perturbation theory in 
section \ref{sec3}. We conclude with some discussion in section \ref{sec4}.

\section{HQET including the $1/m_b$ terms}
\label{sec1} 

We use the Eichten-Hill formulation of HQET \cite{eichten_hill} in which the Lagrangian at order $1/m_b$ is a sum of the leading, 
static, part and two $1/m_b$ corrections
\begin{equation}
\mathscr{L}_{\HQET} = \mathscr{L}_{\stat} - \omega_{\kin} \mathscr{L}_{\kin} - \omega_{\spin} \mathscr{L}_{\spin} 
\end{equation}
with $\mathscr{L}_{\stat} = \bar{\psi}_h D_0 \psi_h$. The power divergent mass-counter term was absorbed 
in ${\color{black} m_{\bare}}$, the only parameter of the static HQET action, which after
an appropriate change of variables appears in a prefactor $e^{-{\color{black}m_{\bare}}|x_0-y_0|}$ of some correlation functions. 

%
The kinetic and chromomagnetic operators enter 
only as insertions in the static vacuum expectation values, namely for some operator $\mathcal{O}$ we have
\begin{equation}
\langle \mathcal{O} \rangle_{\HQET} = \langle \mathcal{O} \rangle_{\textrm{stat}}
+ {\color{black}\omega_{\textrm{kin}}} \sum_{x} \langle \mathcal{O} \mathcal{L}_{\textrm{kin}}(x)
\rangle_{\textrm{stat}}
+ {\color{black}\omega_{\textrm{spin}}} \sum_{x} \langle \mathcal{O} \mathcal{L}_{\textrm{spin}}(x)
\rangle_{\textrm{stat}}.
\end{equation}
Local operators have an effective description as well. We write it explicitely for the lattice discretized 
$A_0$ operator since this will be the operator we will need in the following. We have
\begin{multline}
{\color{black}Z^{\HQET}_{A_0}} (A^{\HQET})_0 = {\color{black}Z^{\HQET}_{A_0}} \Big[ \bar{\psi}_{\ell} \gamma_0 \gamma_5 \psi_h
+ a {\color{black} c_{A_{0,1}}}  \bar{\psi}_{\ell} \frac{1}{2} \gamma_5 \gamma^k \big( \nabla^S_k
- \overleftarrow{\nabla}^S_k \big)\psi_h  +\\+
a {\color{black} c_{A_{0,2}}}  \bar{\psi}_{\ell} \frac{1}{2} \gamma_5 \gamma^k \big( \nabla^S_k
+ \overleftarrow{\nabla}^S_k \big)\psi_h \Big] 
\label{eq. axial current}
\end{multline}
where $\psi_{\ell}$ denote relativistic, massless fermions, whereas $\psi_h$ is a nonrelativistic heavy fermion with $P_+ \psi_h = \psi_h$. 
The renormalization schemes for $Z^{\QCD}_{A_0}$ and $Z^{\HQET}_{A_0}$ will be specified in section \ref{sec. c}. 
Notation for the finite differences $\nabla^S_k$ is taken from \cite{leshouches}.

In order to define HQET and the currents 
at the next-to-leading order in $1/m_b$ one has to fix 3 parameters in $\mathcal{L}_{\HQET}$ 
and 2 $\times$ 3 parameters in $A_0(x)$ and $V_0(x)$ and 2 $\times$ 5 
in $A_k(x)$ and $V_k(x)$ giving in total 19 parameters. They are usualy denoted collectively by $\omega_i$, with $i=1, \dots, 19$. 
In this work we concentrate on two parameters appearing in Eq.\eqref{eq. axial current}, namely
${\color{black} c_{A_{0,2}}} \equiv {\color{black}\omega_5}$ 
and 
${\color{black} Z^{\HQET}_{A_0}} \equiv {\color{black}\omega_6}$ 
and on the corresponding matching observables.


\section{Two examples of matching observables}
\label{sec2}

HQET parameters are determined by considering an appropriately choosen set of observables $\big\{ \Phi_i \big\}_{i=1,\dots,19}$.
The approach implemented by the ALPHA collaboration \cite{3} consists in using the Schr\"odinger functional (SF)
framework \cite{leshouches2} to define correlation functions out of which the observables $\Phi_i$ are constructed.  
In this work we will need one boundary-to-boundary and one boundary-to-bulk correlation 
function, e.g.
\begin{align}
F_1(\theta_{\ell}, \theta_h) &= -\frac{a^{12}}{2L^6} \sum_{\textbf{u}, \textbf{v}, \textbf{y}, \textbf{z}}\langle \bar{\zeta}'_{\ell}(\textbf{u}) \gamma_5
\zeta'_h(\textbf{v}) \bar{\zeta}_h(\textbf{y}) \gamma_5 \zeta_{\ell}(\textbf{z}) \rangle, \\
f_{A_{0}}(\theta_{\ell}, \theta_h, x_0) &= - \frac{a^6}{2} \sum_{\textbf{u}, \textbf{v}} \langle \bar{\zeta}_h(\textbf{u}) \gamma_5 \zeta_{\ell}(\textbf{v})
\big(A_{0}\big)_I(x_0) \rangle 
\end{align}
where $\zeta$ and $\bar{\zeta}$ denote fermionic fields living on the boundary. The $\theta$ angles 
are additional kinematic parameters which in the free theory correspond to the momenta of quark fields,
\begin{equation}
\psi_h(x+ L\hat{k}) = e^{i\theta^k_h} \psi_h(x), \qquad \qquad \psi_{\ell}(x+ L\hat{k}) = e^{i\theta^k_{\ell}} \psi_{\ell}(x).
\end{equation}
The $\theta$ angles can be tuned such as to minimize $1/m_b^2$ effects \cite{3}.
In order to determine ${\color{black} c_{A_{0,2}}}$ and ${\color{black} Z^{\HQET}_{A_0}}$ the following observables were proposed
\begin{align}
\Phi_5(\theta_{\ell}, \theta_{h_1}, \theta_{h_2}) &= \log \frac{f_{A_0}(\theta_{\ell}, \theta_{h_1},x_0=T/2)}{f_{A_0}(\theta_{\ell},\theta_{h_2},x_0=T/2)}, \\
\Phi_6(\theta_{\ell}, \theta_h) &= \log \frac{-Z_{A_0} f_{A_0}(\theta_{\ell}, \theta_h, x_0=T/2)}{\sqrt{F_1(\theta_{\ell}, \theta_h)}} \equiv \log Z_{A_0} + \phi_6(\theta_{\ell}, \theta_h).
\label{obs}
\end{align}
$\Phi_5$ is defined in such a way as to cancel all renormalization factors, whereas in $\Phi_6$ only the renormalization factor of ${A_0}$,
remains uncancelled. A generic matching condition for the '$\Phi_5$-type' observables can be written as
\begin{equation}
\Phi_{i,\QCD}(\bar{m}(m),a = 0,L) \stackrel{!}{=} \Phi_{i,\HQET}(a, L, {\color{black}\omega(\bar{m}(m),a)} )
= \Phi_{i,\stat}(a,L) + \sum_j \Phi_{ij,1/m}(a,L) \ {\color{black}\omega_j}(\bar{m}(m),a),
\label{matching}
\end{equation}
where $L$ is the size of the finite SF volume in which the observables $\Phi_i$ are defined, $a$ is the lattice 
spacing and $\bar{m}(m)$ is the \textbf{b} quark mass defined in the lattice minimal subtraction scheme \cite{leshouches2} 
at the scale $m$ of the \textbf{b} quark mass. The scale $m$ can be given by $m_{\pole}$ or $\bar{m}$ or in any other 
scheme since at one-loop precision the scheme is not relevant. In the following we will work with dimensionless quantities
so we introduce $z$ as a parameter to fix the heavy quark mass 
\begin{equation}
z = \bar{m}(m) L. \label{z}
\end{equation}


The perturbative analysis of the observables Eq.\eqref{obs} was made using \verb[pastor[, an automatic 
tool for generation and calculation of lattice Feynman diagrams \cite{pastor} with SF boundary conditions. 
For a given discretized action, correlation function and parameters such as $L/a$ and the dimensionless 
heavy quark mass $z$, \verb[pastor[ generates the Feynman rules, all Feynman diagrams and a C++ program to evaluate each diagram.
The calculations were performed using the Wilson plaquette gauge action and $\mathcal{O}(a)$-improved Wilson fermions. 

In Ref.\cite{3} a tree-level analysis of the entire set of matching observables was presented. 
The purpose of this work is, using an example of two matching observables, to confirm that $1/m_b^2$ corrections
are small also at one-loop level. Similar results for other observables were reported in \cite{dirk,beauty}.

\section{One-loop contributions to matching observables}
\label{sec3}

\subsection{$c_{A_{0,2}}$}
\label{sec. c}

$f_{A_0}^{\stat}(\theta_l, \theta_h, x_0)$ does not depend on $\theta_h$, therefore $\Phi_{5,\stat}$ vanishes.
We expand the matching condition Eq.\eqref{matching} in $g^2$ and get (abbreviating $(\theta_{\ell}, \theta_{h_1}, \theta_{h_2})$ by $\theta$)
\begin{equation}
{\color{black}\Phi_{5, \QCD}^{(0)}(\theta, z)} + g^2 \Phi_{5, \QCD}^{(1)}(\theta, z) = 
z^{-1} \sum_t \Big( {\color{black}\hat{\omega}_{t}^{(0)} \hat{\Phi}_{5, t}^{(0)}(\theta)} + g^2 \hat{\omega}_{t}^{(1)}(z) \hat{\Phi}_{5, t}^{(0)}(\theta) + 
g^2 \hat{\omega}_{t}^{(0)} \hat{\Phi}_{5,t}^{(1)}(\theta) \Big),
\label{exp}
\end{equation}
with $\hat{\omega}_j = \bar{m} \omega_j$ and $\hat{\Phi}_j = L \Phi_j$.
The sum over $t$ refers to different subleading contributions, namely $t = \{\kin, \spin, c_{A_{0,1}}, c_{A_{0,2}} \}$.
Separating different orders in $g^2$ we get
\begin{align}
{\color{black}\Phi_{5, \QCD}^{(0)}(\theta, z)} &= 
z^{-1} \sum_j {\color{black}\hat{\omega}_{j}^{(0)} \hat{\Phi}_{5, j}^{(0)}(\theta)}, \nonumber \\
\Phi_{5, \QCD}^{(1)}(\theta, z) &= 
z^{-1} \sum_j \Big( \hat{\omega}_{j}^{(1)}(z) \hat{\Phi}_{5, j}^{(0)}(\theta) + 
\hat{\omega}_{j}^{(0)} \hat{\Phi}_{5,j}^{(1)}(\theta) \Big) .
\end{align}
In order to isolate the leading $1/z$ dependence we define a ratio $R$ of the one-loop correction to the tree-level
contribution 
\begin{align}
R_5 = \frac{\Phi_{5, \QCD}^{(1)}(\theta, z)}{{\color{black}\Phi_{5,\QCD}^{(0)}(\theta, z)}} \label{def_r}
&= \frac{ \sum_j {\color{black}\hat{\omega}_{j}^{(0)}} \hat{\Phi}_{5,j}^{(1)}(\theta)}{ \sum_j {\color{black}\hat{\omega}_{j}^{(0)}}
{\color{black} \hat{\Phi}_{5,j}^{(0)}(\theta)}} +
\frac{ \sum_j \hat{\omega}_{j}^{(1)}(z) {\color{black} \hat{\Phi}_{5,j}^{(0)}(\theta)} }{\sum_j {\color{black}\hat{\omega}_{j}^{(0)}
{\color{black} \hat{\Phi}_{5,j}^{(0)}(\theta)}}} 
= {\color{black}\alpha(\theta)  + {\color{black} \gamma(\theta)} \log(z)}  + {\color{black} \mathcal{O}(1/z)}, 
\end{align}
where we used the fact that the only way a $z$-dependence can appear on the right-hand side of the above equation
is through $\hat{\omega}_{j}^{(1)}(z)$ which must be of the functional form 
$\hat{\omega}_{j}^{(1)}(z) = a_j + b_j \log z$ ($a_j$, $b_j$ constants).
When $R$ is plotted on a linear-log plot, it measures:
\begin{itemize}
\item {\color{black} \emph{$1/z^{2}$ corrections}}: deviations from a linear behaviour,
\item {\color{black} \emph{coefficient of the subleading logarithm} }: slope of the data.
\end{itemize}

\begin{figure}
\subfigure{\includegraphics[width=0.495\textwidth]{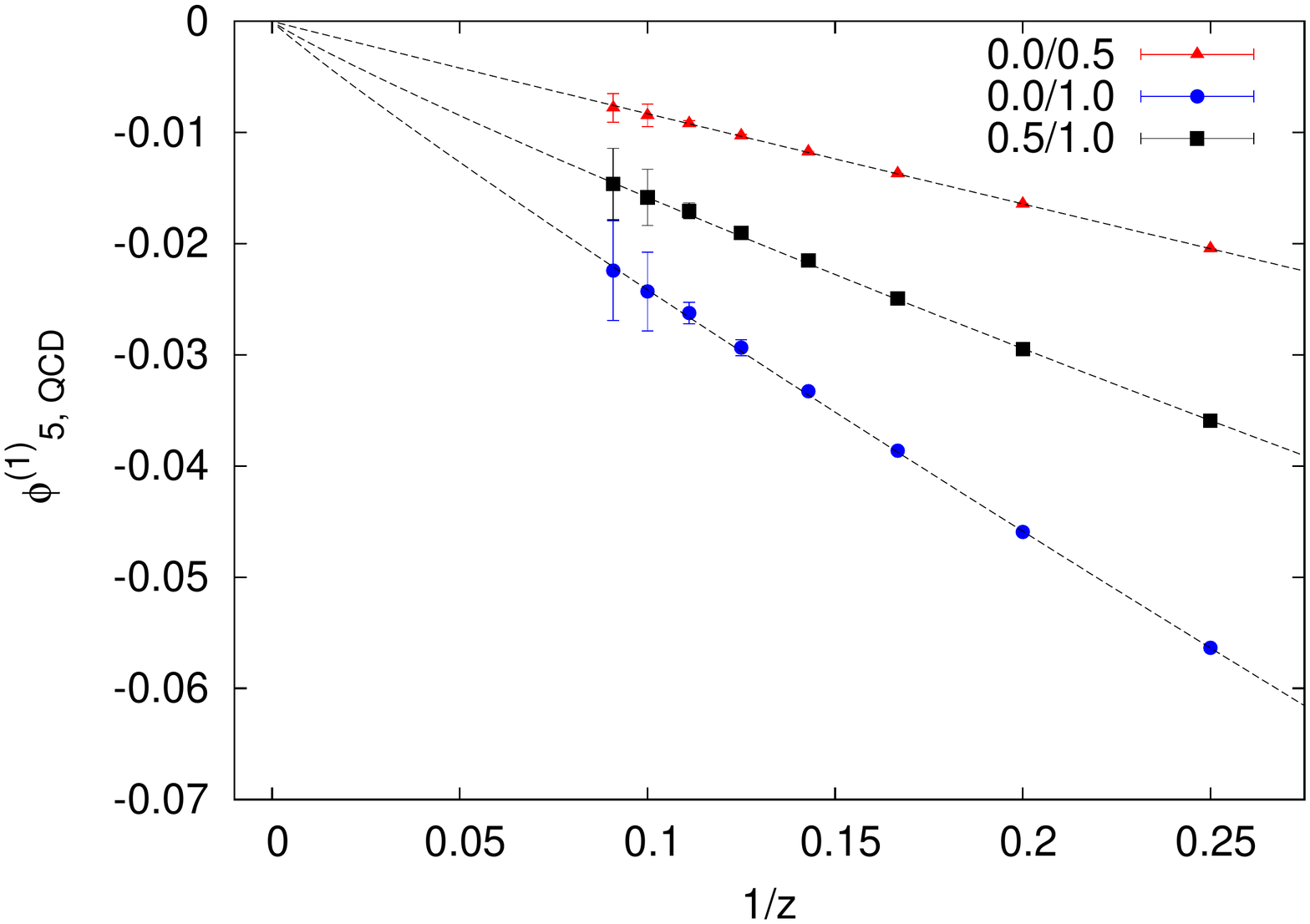} \label{fig. a}}
\subfigure{\includegraphics[width=0.495\textwidth]{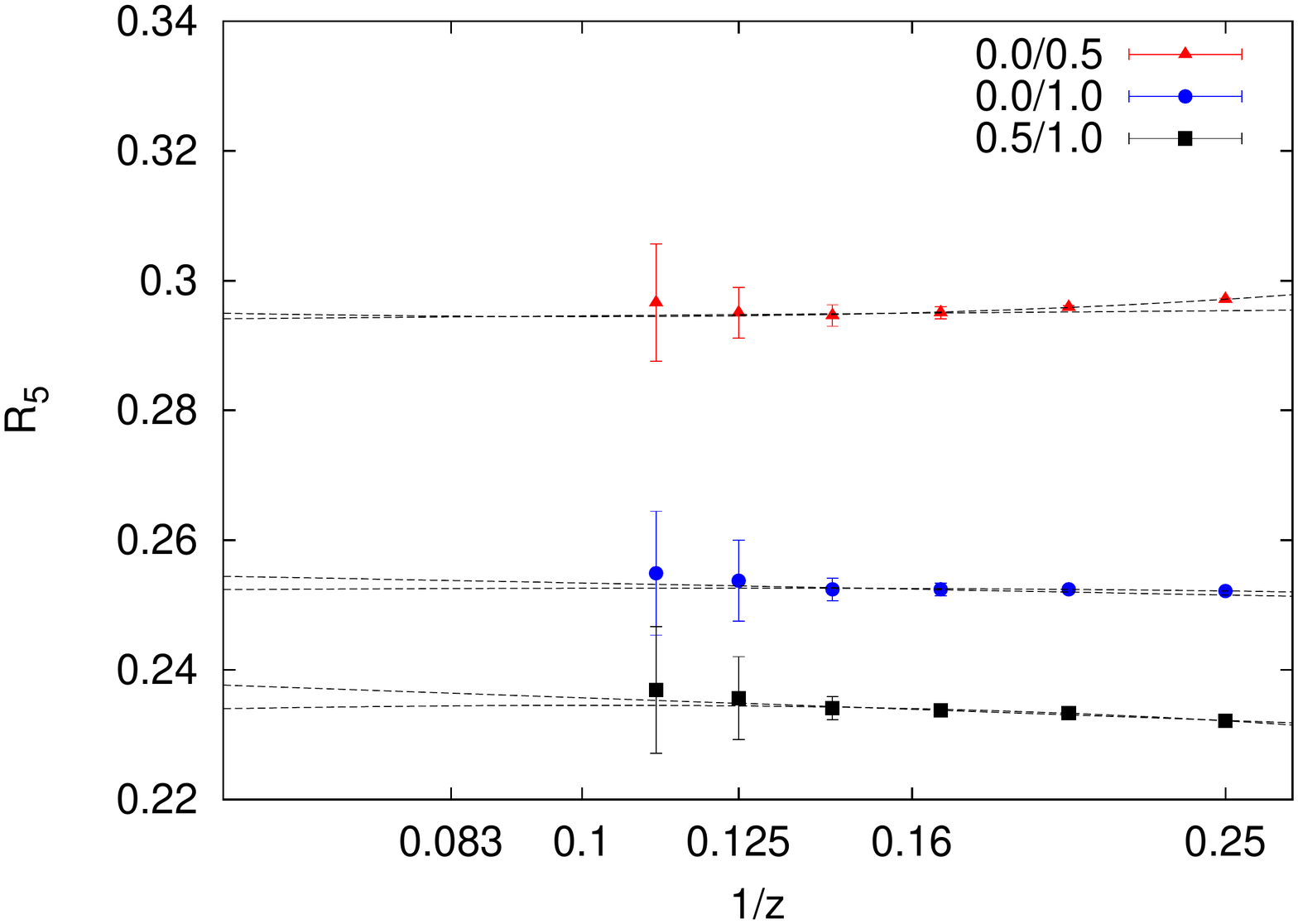} \label{fig. b}}
\caption{
Results for $\Phi_5$. Figure on the left presents the $z$ dependence 
of the one-loop contributions to QCD observables together with a fit of the form $f(z) = \beta_0/z + \beta_1 \log z/z$. 
Figure on the right shows the corresponding $R$ ratio. To each data set two fits were performed with anst\"atze $f(z) = \alpha + \gamma \log z$
 and $f'(z) = \alpha' + \gamma' \log z + \delta'/z$.
One can estimate higher-order corretions by calculating $\frac{f(4) - f'(4)}{f(4)} \sim 0.0003$, which turns out to be very small.
\label{fig. ab}
}
\end{figure}

Plots shown on figure \ref{fig. ab} present the results for the $\Phi_5$ observable. The left plot \ref{fig. a}
shows the one-loop contributions to $\Phi_{5,\QCD}$ extrapolated to the continuum as a function of $z$ which 
extrapolates to a vanishing static limit.
The $1/z^2$ corrections seem to be surprisingly small. The right plot \ref{fig. b} contains 
data for the corresponding $R$ ratio which confirms this observation; the logarithmic dependence as well as 
higher corrections in $1/z$ are very small. Thus, the one-loop results do not favour any of the analyzed 
combination of $\theta$ angles.

\begin{figure}
\subfigure{\includegraphics[width=0.495\textwidth]{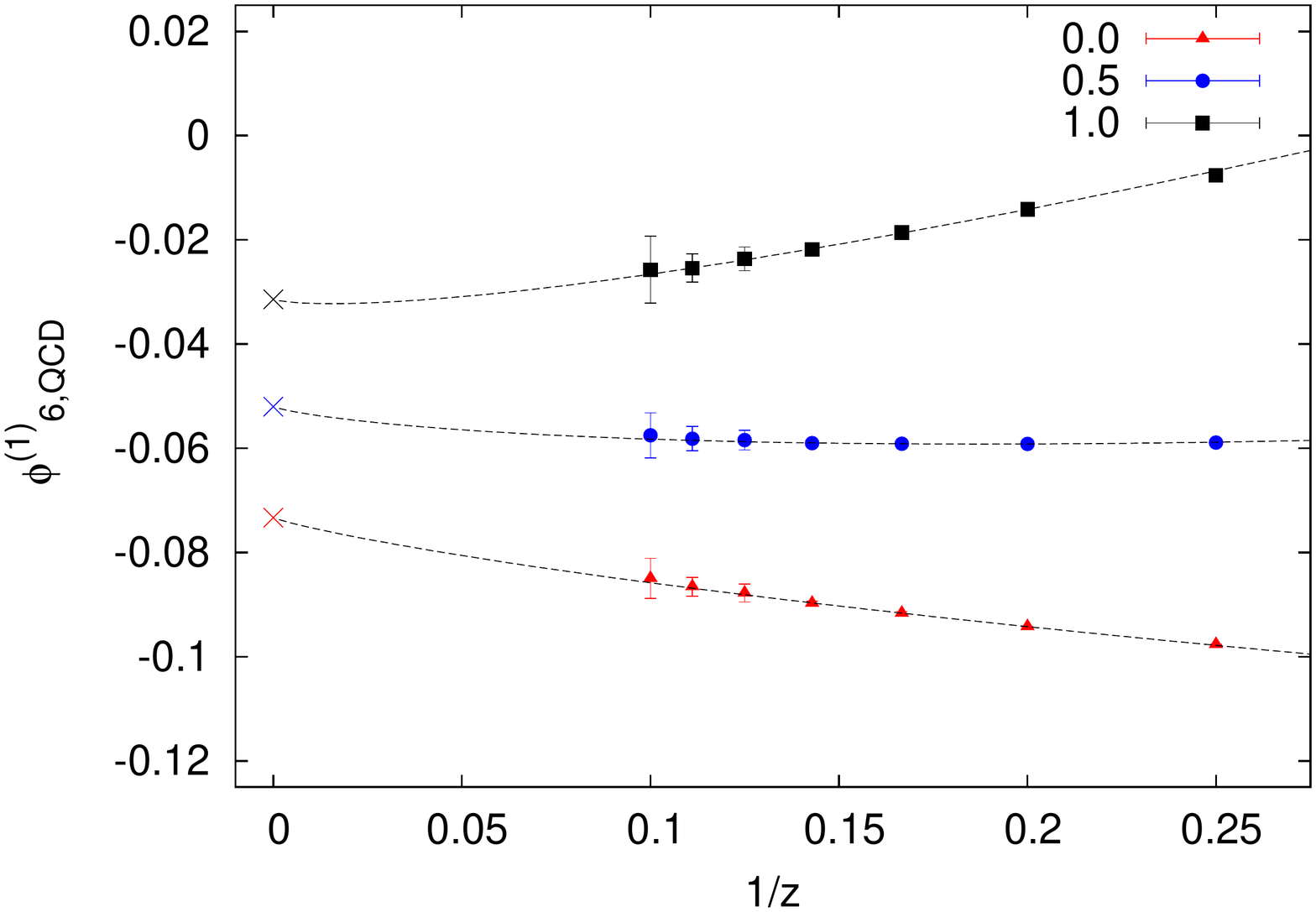} \label{fig. c}}
\subfigure{\includegraphics[width=0.495\textwidth]{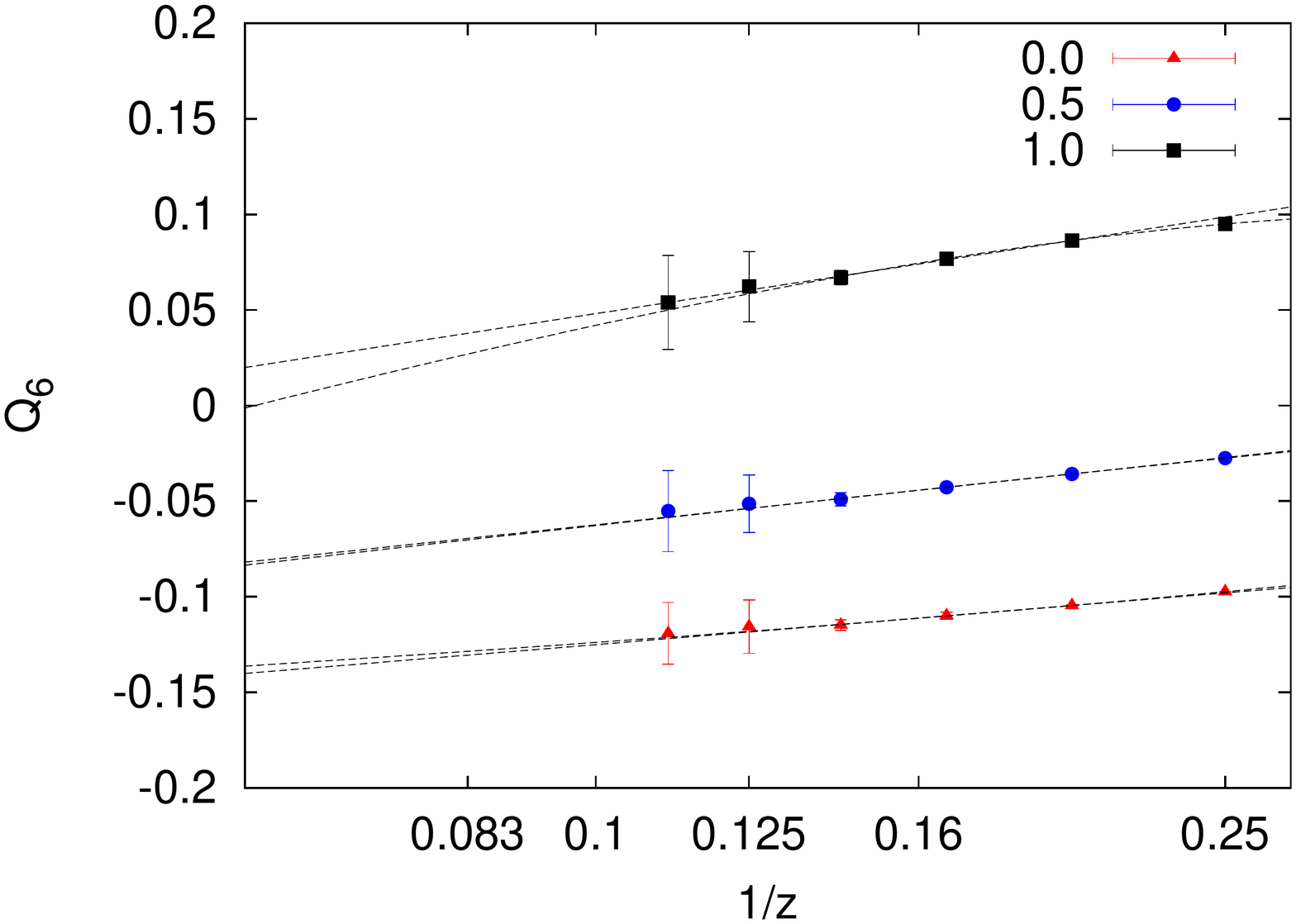} \label{fig. d}}
\caption{
Results for $\Phi_6$. Figure on the left presents the $z$ dependence
of the one-loop contributions to QCD observables together with a fit of the form $f(z) = \beta_0 + \beta_1/z + \beta_2 \log z/z$.
Figure on the right shows the corresponding $R$ ratio. To each data set two fits were performed with anst\"atze 
$f(z) = \alpha + \gamma \log z$  and $f'(z) = \alpha' + \gamma' \log z + \delta'/z$.
One can estimate higher-order corretions by calculating $\frac{f(4) - f'(4)}{f(4)} \sim 0.003$, which turns out to be small.
\label{fig. cd}
}
\end{figure}

\subsection{$Z_{A_0}^{\HQET}$}
\label{sec. z}

In order to fix the renormalization constant $Z_{A_0}^{\HQET}$ we have to match the renormalized observables. 
Writing explicitly the renormalization factors, Eq.\eqref{matching} becomes
\begin{multline}
\lim_{a/L\rightarrow 0} \big[ \log Z^{\QCD}_{A_0} + \phi_{6,\QCD}(z,a/L) \big] \stackrel{!}{=} \log Z_{A_0}^{\HQET}(\mu,a) + \phi_{6,\HQET}(a/L,  {\color{black}\omega}(z,a/L) ) = \\
= \log Z_{A_0}^{\HQET}(\mu, a) + \Big( \phi_{6,\stat}(a/L) + \sum_j \phi_{6,j}(a/L) \ {\color{black}\omega_j}(z,a/L) \Big),
\label{eq. zahqet}
\end{multline}
where the continuum limit is taken keeping the renormalized mass $\bar{m}$ and coupling $g^2$ fixed.
In order to estimate the $1/m_b^2$ corrections to the observable $\phi_6$ it is enough to work at the static
order at which the renormalization factor $Z_{A_0}^{\stat}$ is known. Hence, the matching condition Eq.\eqref{eq. zahqet} 
takes the form 
\begin{equation}
\lim_{a/L \rightarrow 0}\big[ \log Z^{\QCD}_{A_0} + \phi_{6,\QCD}(z,a/L) \big] \stackrel{!}{=} \log C_{A_0}^{\match}  + \log Z_{A_0}^{\stat}(\mu = \bar{m}(m),a)  + \phi_{6,\stat}(a/L) + \mathcal{O}(1/z).
\label{eq. zastat}
\end{equation}
The QCD side is renormalized in a scheme enforcing the current algebra relations at $z=0$ \cite{za}. 
On the HQET side we use an intermediate renormalization scheme, the lattice minimal subtraction scheme, which only cancels
the logarithmic divergence present in $\phi_{6,\stat}(a/L)$ \cite{shifman, politzer}, i.e. 
\begin{equation}
Z_{A_0}^{\stat}(\mu,a) = 1 - \gamma_0 \log(a \mu) g^2 + \mathcal{O}(g^4), \qquad \gamma_0 = -\frac{1}{4\pi^2},
\end{equation}
whereas the finite factor $C^{\match}_{A_0}$ can be used to fix the finite translation factor between the two schemes. 
We explicitly indicated in Eq.\eqref{eq. zastat} that the HQET side was renormalized at 
the scale $\mu = \bar{m}(m)$. In this situation the expansion of the factor $C^{\match}_{A_0}$ is known \cite{sommer_kurth}
\begin{equation}
C^{\match}_{A_0} = 1 + B_{A_0} g^2 + \mathcal{O}(g^4), \qquad B_{A_0} = -0.137(1).
\end{equation}
Eq.\eqref{eq. zastat} can be rewritten as
\begin{equation}
\Phi_{6,\QCD}(z) = \Phi_{6,\stat}(z, a/L) + \log C^{\match}_{A_0}(g^2) + \mathcal{O}(1/z), 
\label{eq. zren}
\end{equation}
We expand both sides of Eq.\eqref{eq. zren} in the coupling $g^2$ and get
\begin{align}
\Phi_{6,\QCD}^{(0)}(z) &= \phi_{6,\stat}^{(0)} + \mathcal{O}(1/z), \nonumber \\
\Phi_{6,\QCD}^{(1)}(z) &= \phi_{6,\stat}^{(1)}(a/L) - \gamma_0 \log (a \bar{m}) + B_{A_0} + \mathcal{O}(1/z). \nonumber
\end{align}
Subtracting $\gamma_0 \log z$ from both sides of the last equation yields
\begin{equation}
\Phi_{6,\QCD}^{(1)}(z) - \gamma_0 \log z  = \phi_{6,\stat}^{(1)}(a/L)
- \gamma_0 \log (a/L) + B_{A_0} + \mathcal{O}(1/z) \equiv \Phi_{6,\stat}^{(1)} + \mathcal{O}(1/z),
\end{equation}
where we consistently used the facts that $\mu = \bar{m}(m)$ and $z=L\bar{m}(m)$.
The sum of subleading terms denoted by $\mathcal{O}(1/z)$ must vanish in the static limit, 
therefore we can assume that at one-loop level it can be parametrized by 
$\alpha_0/z + \alpha_1/z \log z$ ($\alpha_0$, $\alpha_1$ functions of $\theta$ angles only). 
Then, in order to make visible the $1/m_b^2$ corrections we define the quantity $Q$ as
\begin{align}
Q_6 &= z \Big[ \Phi_{6,\QCD}^{(1)}(z) - \gamma_0 \log z \Big] - z \Big[\Phi_{6,\stat}^{(1)} \Big] \nonumber \\
&= z \Big[ \mathcal{O}(1/z) + {\color{black} \mathcal{O}(1/z^2)} \Big] = \alpha_0 + {\color{black} \alpha_1} \log(z) + {\color{black} 
\mathcal{O}(1/z)}
 \nonumber
\end{align}
In analogy to the ratio $R$ of the previous subsection, when $Q$ is plotted on a linear-log plot one can read off
\begin{itemize}
\item the {\color{black} \emph{$1/z^{2}$ corrections}}: as deviations from a linear behaviour,
\item the {\color{black} \emph{coefficient of the subleading logarithm} }: as the slope of the data.
\end{itemize}


Results for the matching observable $\Phi_6$ are presented on figure \ref{fig. cd}. The left plot
\ref{fig. c} shows the z-dependence of the combination $\Phi_{6,\QCD}^{(1)}(z) - \gamma_0 \log z$ together
with the static observable $\Phi_{6,\stat}^{(1)}$. On the right plot \ref{fig. d} 
we show the data for the quantity $Q_6$. Again, the subleading logarithm as well as higher-order corrections in $1/z$ are
small. The one-loop results favour small $\theta$ angles.

\section{Conclusions}
\label{sec4}

In this work we presented a pertubative study of matching observables proposed to match 
non-perturbatively lattice HQET to QCD. We extended the tree-level investigation of Ref.\cite{3} to one-loop order
and discussed in details results for two matching observables. By defining suitable quantities ($R$ and $Q$) we were
able to show that the matching observables do not receive large $1/m_b^2$ corrections at one-loop level, thus
confirming the tree-level conclusions. Complete results for the remaining matching conditions will be presented 
elsewhere \cite{4}.

\acknowledgments
The author would like to thank D. Hesse for the help with \verb[pastor[ and R. Sommer, P. Fritzsch, A. Ramos and H. Simma 
for many useful discussions. The author acknowledges partial financial support by Foundation for Polish Science.

\end{document}